\documentclass[pra,showpacs,preprint]{revtex4}

\usepackage[dvips]{graphicx}

\begin{document}

\title{Finite-size effects in Anderson localization of one-dimensional
  Bose-Einstein condensates}

\author{J. C. C. Cestari, A. Foerster, M. A. Gusm\~ao}
\affiliation{Instituto de F\'isica, Universidade Federal do Rio Grande
  do Sul, Porto Alegre, RS, Brazil}

\begin{abstract}
  We investigate the disorder-induced localization transition in
  Bose-Einstein condensates for the Anderson and Aubry-Andr\'e models
  in the non-interacting limit using exact diagonalization.  We show
  that, in addition to the standard superfluid fraction, other tools
  such as the entanglement and fidelity can provide clear signatures
  of the transition. Interestingly, the fidelity exhibits good
  sensitivity even for small lattices.  Effects of the system size on
  these quantities are analyzed in detail, including the determination
  of a finite-size-scaling law for the critical disorder strength in
  the case of the Anderson model.
\end{abstract}

\maketitle

\section{Introduction} \label{sec:intro}

Localization of waves in a disordered medium was first predicted by
Anderson \cite{Anderson} in the context of non-interacting electrons
in a random crystal.  Recently, developments in the research into
ultracold atomic gases have enlarged the possibilities for studying
Anderson localization in new systems. A breakthrough in the area was
the experimental realization of Anderson localization of a
Bose-Einstein condensate (BEC) in two different kinds of disordered
optical potentials \cite{Billy08,Roati08}: Billy \emph{et al.\/}\
\cite{Billy08} employed a BEC of $\,^{87}$Rb atoms in the presence of a
controlled disorder produced by a laser speckle, while Roati\emph{ et
  al.\/}\ \cite{Roati08} utilized a BEC of essentially non-interacting
$^{39}$K atoms in combination with a one-dimensional quasi-periodic
lattice to observe Anderson localization.

Subsequently, this issue has been attracting increasing interest from
both the experimental and theoretical communities.  From the
experimental point of view, a variety of techniques have been
implemented, such as bichromatic optical lattices
\cite{Roati08,roth,fallani}, speckle laser patterns
\cite{Billy08,Inguscio_speckle,aspect_1d_disorder}, and disordered
cold atom lattices \cite{castin}.  Theoretically, different approaches
such as a variational method \cite{adhikari1}, quantum Monte Carlo
simulation \cite{Haas_QMC,Zimanyi_QMC}, exact diagonalization
\cite{roth,Sengstock_exact,pugatch,Tsubota_exact}, renormalization
group \cite{Singh_RG, Giamarchi_RG}, density-matrix renormalization
group \cite{Zwerger_DMRG, iucci_DMRG, roux_DMRG}, mean-field
approximations
\cite{Buonsante_1d_MFT,Lewenstein_Bose_anderson_glass,Sheshadri,
  dalfovo}, and a perturbative treatment \cite{orso}, among others,
have been explored in this context of disordered ultracold atoms.

The usual model for bosons on a lattice utilizes the so-called
Bose-Hubbard Hamiltonian, with the general form
\begin{equation} \label{ham}
  H = \sum_{i}\varepsilon_i n_i + \Omega\sum_{\langle ij \rangle}(
  a^{\dag}_i a_j +a^{\dag}_j a_i ) + \frac{U}{2} \sum_{i} n_i (n_i -
  1) \,,
\end{equation}
with the standard notation for creation, annihilation, and number
operators for bosons at lattice sites. Each site is considered to have
a single bound state of energy $\varepsilon_i$, the hopping between
sites is restricted to nearest neighbors, with amplitude $\Omega$, and
$U$ is a local repulsive interaction.

Randomness is introduced via the values of $\varepsilon_i$. These
values can have a truly random (usually uniform) distribution in the
range $-\Delta/2 \le \varepsilon_i \le \Delta/2$, which characterizes
the Anderson model \cite{Anderson}. Alternatively, they can vary
periodically with a period incommensurate with the lattice
spacing. This is the case in the Aubry-Andr\'e (AA) model \cite{aa}
for which the energies, in the one-dimensional case, are written as
$\varepsilon_i = \Delta \cos(2\pi\beta i)$, where
$\beta=(1+\sqrt{5})/2$ is the golden ratio, and $i$ assumes integer
values from 1 to $L$ (the system size in units of the lattice
spacing).  Experimentally, speckle patterns can be modeled by the
Anderson model, while quasi-periodic potentials can be described by
the AA model.

The phenomenology of model (\ref{ham}) at low temperatures involves
Bose-Einstein condensation, superfluidity, a Mott (gapped) phase, and
Anderson localization, depending on the relative values of $\Omega$,
$U$, and $\Delta$ \cite{Fisher89}. Theoretical studies of such a rich
phenomenology are faced with the usual challenges of interacting
many-body problems.  For arbitrary interaction strengths perturbation
theory cannot be used, which more or less restricts the approach to
numerical solutions on finite-size lattices. Then, finite-size effects
tend to smear out the sharp phase boundaries expected in the
thermodynamic limit. This is true even for the pure Anderson
transition in the non-interacting limit. For instance, in his original
paper \cite{Anderson} Anderson proved that his model presents
localization for any $\Delta > 0$ in one dimension, while Aubry and
Andr\'e \cite{aa} proved that in their model localization only occurs
for $\Delta \ge 2$, but these critical values are not clearly seen in
small lattices.

Our aim in this paper is to discuss in detail the effects of finite
size in the non-interacting limit for a one-dimensional lattice. We
focus on the size dependence of quantities that provide signatures of
quantum phase transitions (QPTs), here applied to Anderson
localization. In addition to the superfluid fraction, common in
studies of Bose-Einstein condensates, we also employ quantities
borrowed from quantum information theory, like entanglement and
fidelity \cite{nielsen}.  All of these quantities allow us to obtain
the correct thermodynamic limit of the critical disorder strength for
both Anderson and Aubry-Andr\'e models. In particular, we are able to
show that the critical disorder strength in the Anderson model obeys a
scaling law with the system size.

\section{Localization transition} \label{sec:loc}

We consider one-dimensional lattices of $L$ sites, and perform exact
numerical diagonalization of the Hamiltonian matrix, obtaining a
chosen number of lowest-energy eigenvalues, and the corresponding
eigenvectors. We will focus only on ground-state results. Since
interaction is not taken into account here, we restrict our numerical
calculations to a single particle. When evaluating the quantities of
interest for the Anderson model, it is necessary to average over a
large number of random configurations of the local energies in order
to have physically meaningful results.

For each lattice size, we calculate the superfluid fraction,
ground-state entanglement and fidelity, as discussed in detail below.
This is done for a wide range of disorder strength values, with the
aim of determining a critical value above which the states are
localized. Throughout the paper, energy quantities (like $\Delta$) are
measured in units of the tunneling amplitude $\Omega$, while the
lattice size $L$ is expressed in units of the lattice spacing, which
is equivalent to the number of sites.

\subsection{Superfluid fraction} \label{ssec:sf}

The ground state of a uniform Bose-Einstein condensate should be
superfluid. There is some controversy as to whether this remains true
in the non-interacting limit. As discussed by Lieb \emph{et al.\/}\
\cite{lieb}, if one defines a superfluid as a fluid that does not
respond to an external velocity field (e.g., rotation of the container
walls), then the ideal Bose gas is a perfect superfluid in its ground
state. This is the approach we adopt here. On the other hand, the
presence of disorder tends to alter the nature of the quantum states,
which become no longer extended over the entire system, but localized
over a finite distance. Then, the so called \emph{superfluid
  fraction\/} (here denoted by $f_s$) can signal whether the states
are extended ($f_s \ne 0$) or not ($f_s = 0$).

The superfluid fraction for a discrete lattice system is proportional
to the difference in ground-state energy between the systems with
periodic and with twisted boundary conditions \cite{roth}. Instead, we
may keep periodic boundary conditions and use a twisted Hamiltonian
\begin{equation}
  H_\theta = \sum_{i=1}^{L}\varepsilon_i n_i + \Omega\sum_{\langle ij
    \rangle}( a^{\dag}_i a_j e^{-i\frac{\theta}{L}} +a^{\dag}_j a_i
  e^{i\frac{\theta}{L}}) \,,
\label{twist}
\end{equation}
where $\theta$ is the twist angle.  Ground-state energies of the
Hamiltonian (\ref{twist}) are calculated for $\theta=0$ ($E_0$) as
well as for finite $\theta$ ($E_\theta$), and the superfluid fraction
is given by \cite{roth}
\begin{equation}
f_s = \frac{L^2}{N\Omega}\frac{(E_\theta-E_0)}{\theta^2} \,,
\label{defsff}
\end{equation}
being independent of the twist angle $\theta$ as long as $\theta <<
\pi$. In practice, $f_s$ remains the same even for twist angles that
are a significant fraction of $\pi$. Typically, we use $\theta = 0.1$
in our numerical calculations.

\begin{figure}[t]
  \begin{center}
  \includegraphics[width=8.6cm]{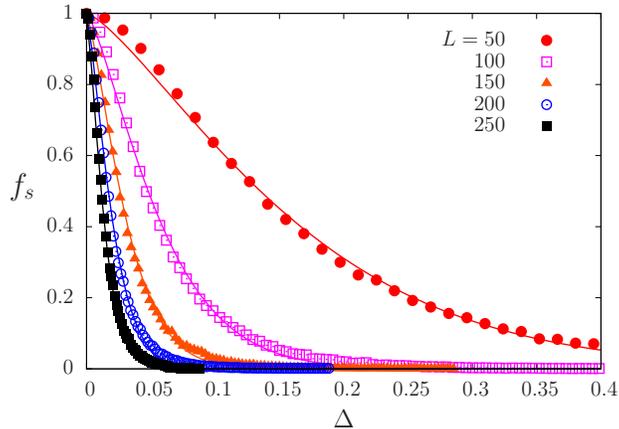}
\end{center}
\caption{(Color online) Average superfluid fraction in the Anderson
  model as a function of the disorder strength $\Delta$ for different
  lattice sizes. The averages were calculated from 5000 random
  configurations of the potential. The lines correspond to fittings
  with Eq.~(\ref{fitsff}).}
  \label{fig:sfAnd}
\end{figure}

In Fig.~\ref{fig:sfAnd} we show the superfluid fraction for the
Anderson model as a function of disorder strength for different
lattice sizes. When $\Delta = 0$ we always have $f_s = 1$, i.e., the
system is in a superfluid phase. For large enough $\Delta$ we have
$f_s \to 0$, and the ground-state is localized.

We define a characteristic value $\Delta_L$ of the disorder strength
for each lattice size by observing that the superfluid-fraction curves
are well fitted by the function
\begin{equation} \label{fitsff}
f_s \simeq e^{(-\frac{\Delta}{\Delta_L})^\alpha} \,.
\end{equation}
For all lattice sizes $L$ we find that the fitting parameter $\alpha$
has approximately the same value $\alpha \simeq \frac{4}{3}$. Lines in
Figure~\ref{fig:sfAnd} show the fittings with Eq.~(\ref{fitsff}) while
symbols are the corresponding values obtained by the numerical
solution.  Fig.~\ref{fig:sfAnd} seems to indicate that $\Delta_L \to
0$ when $L \to \infty$, as expected. This behavior will be verified in
Sec.~\ref{sec:fsizescal} through a finite-size-scaling analysis.

\begin{figure}[t]
  \begin{center}
  \includegraphics[width=8.6cm,clip]{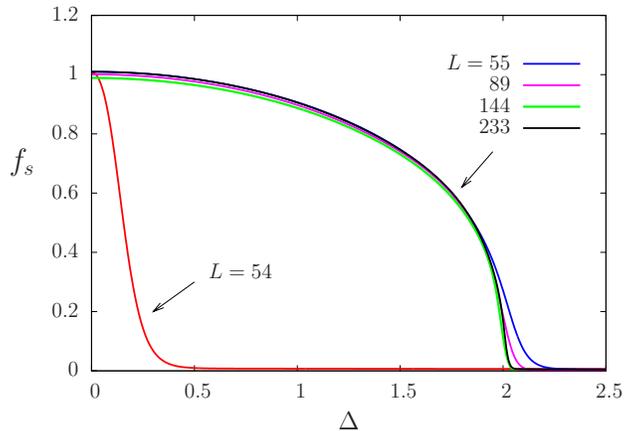}
\end{center}
\caption{(Color online) Superfluid fraction in the AA model as a
  function of the potential amplitude $\Delta$ for chosen lattice
  sizes. The expected critical value $\Delta_c = 2$ is well verified
  for suitably chosen lattice sizes, but not for others (e.g., $L=54$
  in this figure), as discussed in the text. The lines indicated at
  the top right corner are essentially indistinguishable, except near
  $\Delta=2$, where rounding is more noticeable for smaller sizes.}
  \label{fig:sfAA}
\end{figure}

Figure~\ref{fig:sfAA} shows the superfluid fraction for the
Aubry-Andr\'e model as a function of the ``disorder strength''
$\Delta$ for different lattice sizes. Actually, $\Delta$ is the
amplitude of the \emph{Aubry-Andr\'e potential\/}, i.e., the
incommensurate periodic modulation of local energies. We can see that
$f_s(\Delta) = 0$ for $\Delta$ larger than a characteristic
$\Delta_L$, with $\Delta_L \approx 2$, consistent with the expected
critical value $\Delta_c = 2$. However, this happens for certain
lattice sizes, but for other values of $L$ the behavior of $f_s$ with
$\Delta$ is completely different, even for very similar sizes, as
exemplified for $L=54$ and 55 in Fig.~\ref{fig:sfAA}. The reason for
this apparently puzzling feature lies in the mismatching between
periodic boundary conditions with integer period $L$ and the
Aubry-Andr\'e potential with irrational period $\beta^{-1}$. One can
verify that $\Delta_L \approx 2$ occurs when $L$ belongs to the
sequence of Fibonacci numbers \cite{harper}. The first members of this
sequence, for $L \ge 8$, are 8, 13, 21, 34, 55, 89, 144,\ldots. The
ratio between two consecutive Fibonacci numbers approaches the golden
ratio when these numbers become very large. For small lattice sizes it
is better to redefine the parameter $\beta$ to be such a ratio of
Fibonacci numbers. We do this in order to get better results for
smaller sizes like $L=8$ or 13, which become relevant for numerical
solutions in the presence of interactions. It is worth mentioning
that departure from the critical $\Delta_c = 2$ was observed in the
experiments of Roati \emph{et al.\/} \cite{Roati08}. They find
localization for $\Delta \ge 7$, using an incommensurate potential
with $\beta \simeq 1.1972$. This value of $\beta$, being significantly
different from a ratio of consecutive Fibonacci numbers (the golden
ratio is $\sim 1.6180$), does not ensure the duality condition
\cite{harper} that yields $\Delta_c = 2$.

\subsection{Entanglement} \label{ssec:ent}

\begin{figure}[t]
  \begin{center}
  \includegraphics[width=8.2cm,clip]{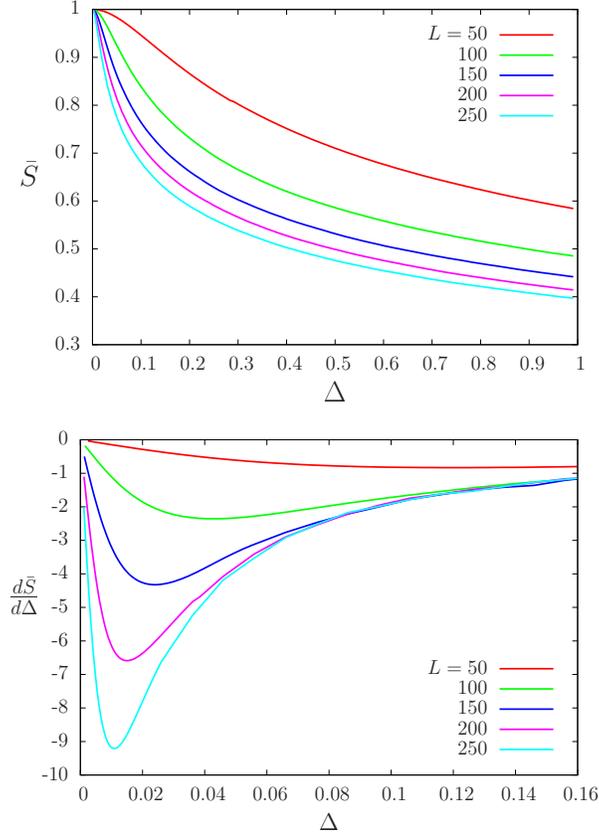}
\end{center}
\caption{(Color online) Top: Average entanglement as a function of the
  disorder strength $\Delta$ for different lattice sizes in the
  Anderson model. Averages were calculated from 5000 random
  configurations of the potential. Bottom: Derivative of the
  entanglement with respect to the disorder strength $\Delta$ for the
  same sizes. The location of its minimum for a given system size $L$
  is taken as the corresponding critical disorder strength
  $\Delta_{L}$. On both panels, curves are associated to increasing
  lattice sizes from top to bottom.}
  \label{fig:entAnd}
\end{figure}

The concept of quantum entanglement can be used to study localization
if we choose a basis for the Hilbert space composed of states that are
localized at each of the lattice sites. Then, a fully extended
ground-state will have maximum entanglement of the basis states. In
the opposite limit, the entanglement will be zero for localization at
a single site. So, the ground-state Shannon entropy is a good measure
of entanglement. It is given by
\begin{equation}
  S =  - \sum_i p_i \log_2 p_i\,,
\label{shan}
\end{equation}
where $p_i=|c_i|^2$, with $c_i$ representing the coefficient of the
$i$-th basis state in the expansion of the ground-state vector
$|\Psi_0\rangle$ \cite{stephan,milburn}. For a lattice of $L$ sites,
the maximum value of $S$ is $\log_2 L$, and occurs when all the $p_i$
are equal. We define $\bar S \equiv S/\log_2 L$ as our measure of
entanglement, whose maximum value will be $\bar S_\mathrm{max} = 1$.

\begin{figure}[t]
  \begin{center}
  \includegraphics[width=8.5cm]{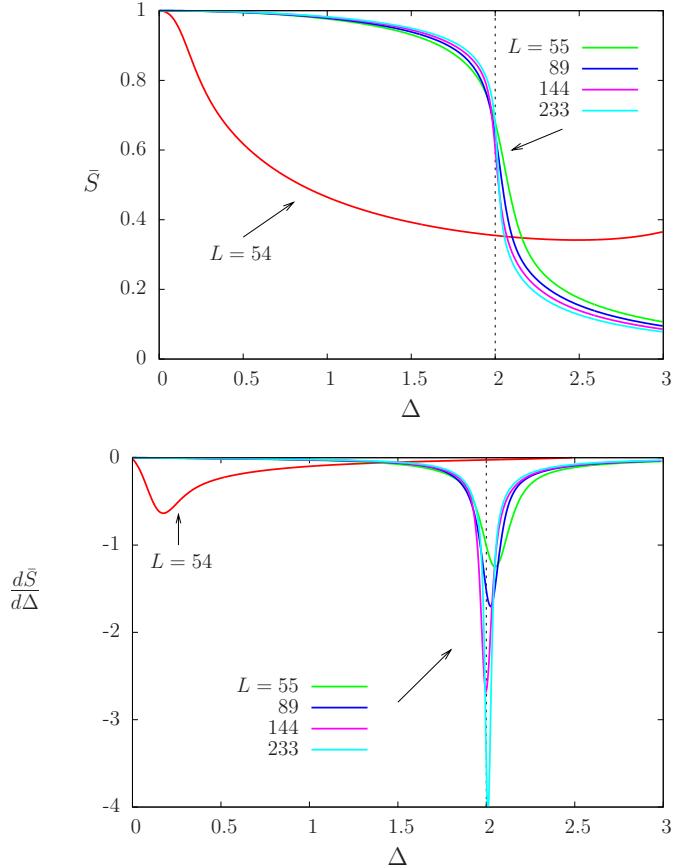}
\end{center}
\caption{(Color online) Entanglement (top) as a function of the
  potential amplitude $\Delta$ for chosen lattice sizes in the AA
  model, and the corresponding derivative (bottom). Notice the sudden
  drop in the first case, and sharp minima in the second near the
  point $\Delta = 2$, except for sizes that are not Fibonacci numbers,
  here exemplified by $L=54$. The size effect is better noticed in the
  bottom panel, where deeper minima correspond to larger lattice
  sizes.}  \label{fig:entAA}
\end{figure}

In Fig.~\ref{fig:entAnd} (top panel) we show the ground-state
entanglement $\bar S$ (averaged over disorder) for the Anderson model
as a function of the disorder strength $\Delta$ for various lattice
sizes $L$. We define the characteristic $\Delta_L$ in this case as the
inflection point of the entanglement curve. It can be better
visualized as the minimum in the derivative of $\bar S$ with respect
to $\Delta$, as shown in the bottom panel of
Fig.~\ref{fig:entAnd}. Notice that the minimum moves toward
$\Delta=0$ as $L$ increases. We will come back to this point in our
finite-size-scaling analysis of Sec.~\ref{sec:fsizescal}.

In the case of the AA model, the ground-state entanglement, when
calculated for lattice sizes belonging to the Fibonacci series, shows
a sudden drop near $\Delta=2$, as can be seen in Fig.~\ref{fig:entAA}
(top panel). This indicates that the derivative $d\bar S/d\Delta$ is
again a good marker of the quantum phase transition. The bottom panel
of Fig.~\ref{fig:entAA} shows this derivative as a function of
$\Delta$ for different lattice sizes. We can see a sharp minimum in
the derivative essentially at $\Delta = 2$, more or less independently
of the lattice size. Here, the size effect appears mainly in the depth
of the minimum, which increases dramatically with increasing size. A
non Fibonacci size has been included for comparison in
Fig.~\ref{fig:entAA}.

\begin{figure}[t]
  \begin{center}
  \includegraphics[width=8.4cm]{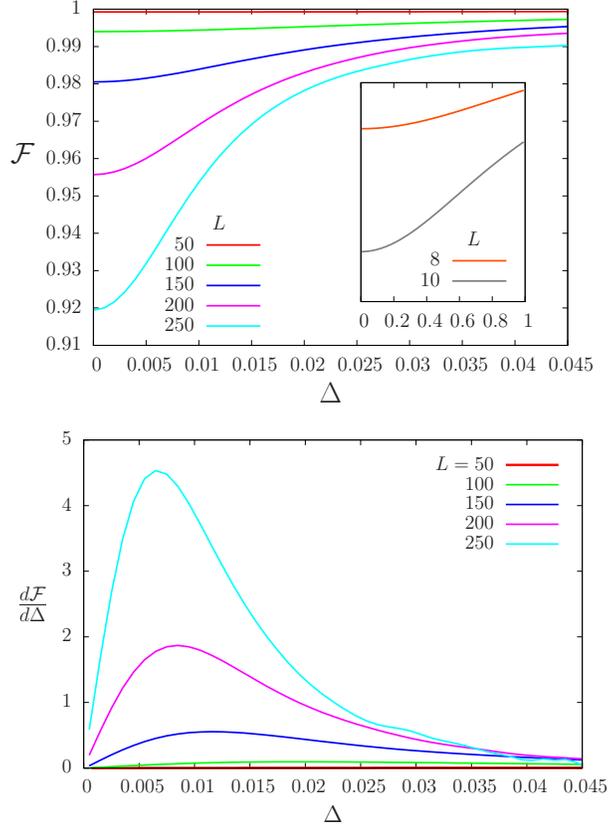}
\end{center}
\caption{(Color online) Average ground-state fidelity (top) in the
  Anderson model as a function of disorder strength for different
  lattice sizes, and the corresponding derivative (bottom). Averages
  were calculated from 5000 random configurations of the
  potential. The top-panel inset shows (in a much smaller vertical
  scale) that the minimum at $\Delta=0$ exists even for small lattice
  sizes. The position of the maximum derivative for a given system
  size $L$ is taken as the critical disorder strength
  $\Delta_{L}$. Deeper minima of the fidelity (higher maxima of its
  derivative) correspond to larger lattice sizes.}
  \label{fig:fidAnd}
\end{figure}

\subsection{Fidelity} \label{ssec:fid}

The \emph{fidelity\/} is a measure of how distinguishable two quantum
states are. For pure states, it is defined as the absolute value of
the overlap between these two states. We use here the scalar product
of two ground-state vectors of Hamiltonians with slightly different
values of $\Delta$, writing the fidelity as
\begin{equation}
\mathcal{F}(\Delta) = \Bigl| \langle \Psi_0(\Delta - \delta\Delta)
| \Psi_0(\Delta + \delta\Delta) \rangle \Bigr|  \, ,
\label{fid}
\end{equation}
which has values between $0$ and $1$.  Generically, the fidelity
exhibits a minimum at a critical parameter value characterizing a QPT.
We have checked that although the choice of $\delta \Delta$ affects
the magnitude of the minimum, which can be made arbitrarily small, the
value of $\Delta$ at which the minimum occurs is independent of
$\delta\Delta$. Here we use $\delta\Delta = 10^{-2}$ in all numerical
calculations. In the case of the Anderson model, we perform the
configuration average on $\mathcal{F}(\Delta)$, which means that both
ground states are obtained for the same realization of the random
local energies.

\begin{figure}[t]
  \begin{center}
  \includegraphics[width=8.6cm,clip]{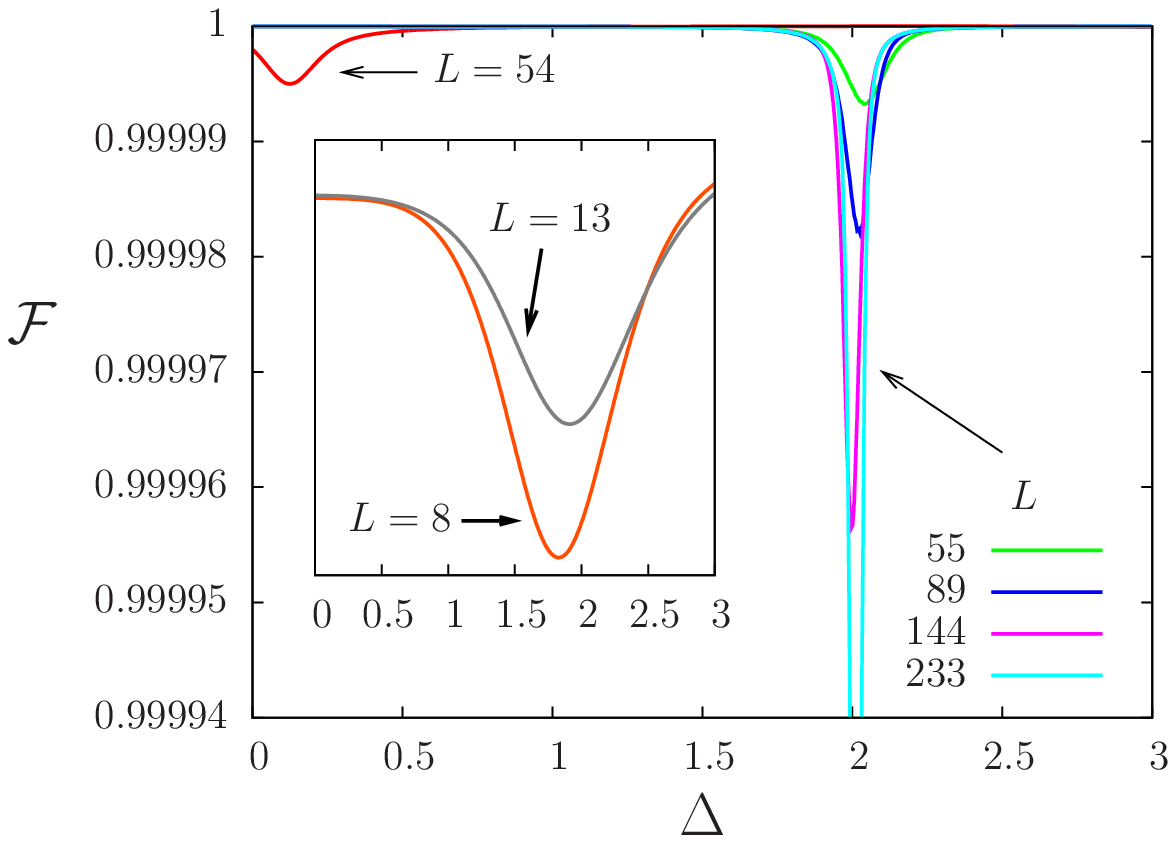}
\end{center}
\caption{(Color online) Fidelity as a function of the disorder
  strength $\Delta$ for different lattice sizes. There is a sharp peak
  close to the point $\Delta = 2$. The inset shows two small lattice
  sizes in a much narrower scale close to $\mathcal{F}=1$. We
  redefined $\beta$ as 13/8 and 21/13 for $L = 8$ and 13,
  respectively. The curves indicated at the bottom right show minima
  that become more pronounced as $L$ increases.} \label{fig:fidAA}
\end{figure}

The average ground-state fidelity for the Anderson model is shown in
Fig.~\ref{fig:fidAnd} (top panel) as a function of the disorder
strength for different lattice sizes. It is interesting to notice that
it has a minimum for all lattice sizes at $\Delta=0$, which is the
expected critical disorder strength in one dimension for $L \to
\infty$. The inset of Fig.~\ref{fig:fidAnd} shows that this remains
true even for sizes as small as $L=8$ and 10. The finite-size effect
is noticeable in the minimum depth, which increases with
$L$. Nevertheless, the minima are relatively broad, and an inflection
point exists, which moves down in $\Delta$ as $L$ increases. This is
better seen in the derivative of $\mathcal{F}(\Delta)$, shown in the
bottom panel of Fig.~\ref{fig:fidAnd}. Similarly to what we observed
for the superfluid fraction, here the maximum of
$d\mathcal{F}/d\Delta$ locates a characteristic disorder strength
$\Delta_{L}$ for each lattice size. A detailed analysis of the
behavior of $\Delta_{L}$ for large $L$ will be presented in
Sec.~\ref{sec:fsizescal}.

The fact that $\mathcal{F}(\Delta)$ can signal the critical disorder
strength for all lattice sizes is confirmed for the AA model.  Figure
\ref{fig:fidAA} shows the fidelity as a function of $\Delta$ for
different lattice sizes (all belonging to the Fibonacci series, except
$L=54$). A sharp minimum appears very close to $\Delta = 2$ for all
the appropriate sizes, clearly indicating the localization
transition. In this case, one does not extract more information from
the derivative. The size effect is evident in the minimum depth, which
is strongly dependent on the value of $L$.

We can observe that the ground-state fidelity is a very precise tool
to indicate the localization transition, even for small lattice
sizes. This is specially relevant in the interacting case, where
numerical solutions must be restricted to relatively small sizes. It
is interesting to mention that the concept of fidelity was also
efficiently applied to identify quantum phase transitions in other
models in the BEC scenario \cite{a1,a2}. For the Bose-Hubbard model
without disorder, it was shown \cite{Buonsante_1d_MFT} that the
fidelity is a clearer indicator of a QPT than the entanglement.

\section{Finite-size scaling} \label{sec:fsizescal}

In the previous section, we have seen that good signatures of the
localization transition in the one-dimensional AA model are provided
by the superfluid fraction, the ground-state entanglement, and the
ground-state fidelity. All these quantities (or appropriate
derivatives) correctly locate the critical amplitude of the
incommensurate potential (equivalent to a ``disorder
strength''), more or less independently of lattice size, provided this
size is not too small, and is chosen to be nearly commensurate with an
integer number of oscillations of the AA potential. 

\begin{figure}[t]
 \begin{center}
  \includegraphics[width=8.8cm,clip]{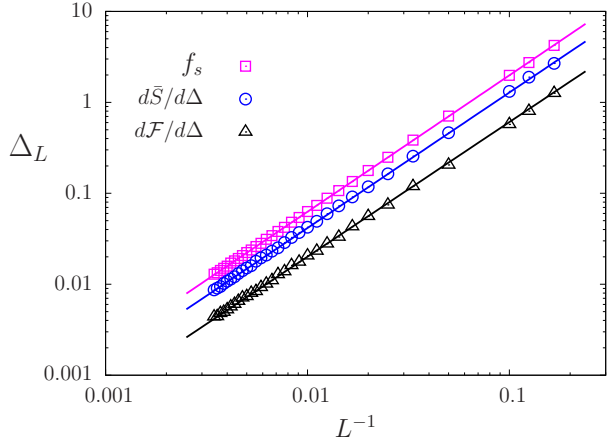}
\end{center}
\caption{(Color online) Logarithmic plot of the critical disorder
  strength $\Delta_L$ in the Anderson model as a function of the
  lattice size, evaluated from superfluid fraction (squares),
  derivative of the entanglement (circles), and derivative of the
  fidelity(triangles). The lines are fittings to the scaling law
  (\ref{scale}), which is satisfied with an exponent $\gamma \approx
  1.5$ and with $\Delta_\infty = 0$ for all three
  quantities. \label{fig:scaling}}
\end{figure}

For the true Anderson model, however, the results are less clear-cut
upon a simple direct visualization. We defined characteristic disorder
strengths $\Delta_L$ in Sec.~\ref{sec:loc}, with different definitions
depending on the physical quantity evaluated.  In order to determine
more quantitatively their size dependence, these different
$\Delta_L$'s where calculated for lattice sites from 6 to 300,
with random averages being performed typically over 5000
configurations. Despite the different definitions, if we make a
log-log plot of $\Delta_L$ vs\ $L^{-1}$, as is done in
Fig.~\ref{fig:scaling}, we see that in all the three cases (superfluid
fraction, entanglement, and fidelity) the following scaling law is
obeyed:
\begin{equation} \label{scale}
\Delta_{L} - \Delta_{\infty} = C\, L^{-\gamma} \,,
\end{equation}
where $C$ and $\gamma$ are positive real numbers, and
$\Delta_{\infty}$ is the critical disorder strength in the
thermodynamic limit. Despite the fact that the
prefactor $C$ in Eq.~(\ref{scale}) differs for the three definitions
of $\Delta_L$, we find the same exponent $\gamma \approx 1.5$ in all
three cases, and the results are consistent with $\Delta_\infty = 0$,
in accordance with the exact value for Anderson localization in one
dimension. It is also worth noticing in Fig.~\ref{fig:scaling} that
the scaling law (\ref{scale}) is satisfied even for lattice sizes as
small as $L = 6$.

\section{Conclusions}

We presented a detailed numerical study of disorder-induced
localization in the one-dimensional Anderson and Aubry-Andr\'e models
of Bose-Einstein condensates. Through exact numerical diagonalization
of the Hamiltonian matrix, we studied the superfluid fraction,
entanglement, and fidelity as quantities that can signal the
localization transition even at finite size. Here we restricted our
analysis to the non-interacting limit in order to reach large lattice
sizes. With this, we were able to verify that finite-size-scaling laws
are obeyed by all the studied quantities in the case of the Anderson
model, with a common exponent. In contrast, finite-size effects in the
Aubry-Andr\'e model are less pronounced, and mainly related to the
incommensurability between the potential and the system size. If the
incommensurate potential is tuned so that an integer number of periods
nearly coincides with the system size, then the localization
transition is fairly sharp at the critical potential strength.

Even though all the three quantities analyzed here show clear
signatures of the localization transition, we find that the fidelity
is less prone to finite-size effects. This makes it specially suited
to be used in the interacting case, where numerical solutions are
restricted, for practical reasons, to relatively small sizes.

\subsection*{Acknowledgments}

We would like to thank CNPq - Conselho Nacional de Desenvolvimento
Cient\'ifico e Tecnol\'ogico (Brazil) for financial support.


\begin{thebibliography}{99}

\bibitem{Anderson} P. W. Anderson, Phys. Rev. \textbf{109}, 1492 (1958).

\bibitem{Billy08} J. Billy, V. Josse, Z. Zuo, A. Bernard,
  B. Hambrecht, P. Lugan, D. Cl\'ement, L. Sanchez-Palencia, P. Bouyer
  and A. Aspect, Nature \textbf{453}, 891 (2008).

\bibitem{Roati08} G. Roati, C. D'Errico, L. Fallani, M. Fattori,
  C. Fort, M. Zaccanti, G. Modugno, M. Modugno, M. Inguscio, Nature
  \textbf{453}, 895 (2008).

\bibitem{roth}
R. Roth and K. Burnett, Phys. Rev. A \textbf{68}, 023604 (2003).

\bibitem{fallani} 
L. Fallani,  J. E. Lye, V. Guarrera, C. Fort, and M. Inguscio,
Phys. Rev. Lett. \textbf{98}, 130404 (2007).

\bibitem{Inguscio_speckle}
J. E. Lye, L. Fallani, M. Modugno, D. S. Wiersma, C. Fort and M. Inguscio,
Phys. Rev. Lett. \textbf{95}, 070401 (2005).

\bibitem{aspect_1d_disorder}
D. Cl\'{e}ment, P. Bouyer, A. Aspect, L. Sanchez-Palencia,
Phys. Rev. A \textbf{77}, 033631 (2008).

\bibitem{castin}U. Gavish and Y. Castin, Phys. Rev. Lett. 
{\bf95}, 020401 (2005).

\bibitem{adhikari1} Y. Cheng and S. K. Adhikari, 
Phys. Rev. A \textbf{82}, 013631 (2010).

\bibitem{Haas_QMC} P. Sengupta, A. Raghavan, and S. Haas, New J. Phys.
  \textbf{9}, 103 (2007).

\bibitem{Zimanyi_QMC}
R. T. Scalettar, G. G. Batrouni and G. T. Zimanyi, 
Phys. Rev. Lett. \textbf{66}, 3144 (1991).   

\bibitem{Sengstock_exact}
D.-S. L\"uhmann, K. Bongs, K. Sengstock, and D. Pfannkuche, 
Phys. Rev. A \textbf{77}, 023620 (2008).

\bibitem{pugatch}
R. Pugatch, N. Bar-Gill, N. Katz, E. Rowen and N. Davidson, 
cond-mat/0603571v6.

\bibitem{Tsubota_exact}
P. Louis, M. Tsubota, J. Low Temp. Phys. \textbf{148}, 351 (2007).

\bibitem{Singh_RG}
K.G. Singh and D.S. Rokhsar,  Phys. Rev. B \textbf{46}, 3002 (1992). 

\bibitem{Giamarchi_RG}
T. Giamarchi and H. J. Schulz, Phys. Rev. B \textbf{37}, 325 (1988).

\bibitem{Zwerger_DMRG} S. Rapsch, U. Scholho\"ock, and W. Zwerger,
  Europhys. Lett. \textbf{46}, 559 (1999).

\bibitem{iucci_DMRG}
C. Kollath, A. Iucci, T. Giamarchi, W. Hofstetter and U. Schollw\"ock, 
Phys. Rev. Lett. \textbf{97}, 050402 (2006).

\bibitem{roux_DMRG} 
G. Roux, T. Barthel, I. P. McCulloch, C. Kollath, U. Schollw\"ock, and
T. Giamarchi, Phys. Rev. A \textbf{78}, 023628 (2008).

\bibitem{Buonsante_1d_MFT}				
P. Buonsante, V. Penna, A. Vezzani and P. B. Blakie,
Phys. Rev. A \textbf{76}, 011602(R) (2007). 

\bibitem{Lewenstein_Bose_anderson_glass}		
B. Damski, J. Zakrzewski, L. Santos, P. Zoller and M. Lewenstein,
Phys. Rev. Lett. \textbf{91}, 080403 (2003). 

\bibitem{Sheshadri}	
K. Sheshadri, H. R. Krishnamurthy, R. Pandit and T. V. Ramakrishnan, 
Phys. Rev. Lett. \textbf{75}, 4075 (1995).

\bibitem{dalfovo} M. Larcher, F. Dalfovo and M. Modugno, Phys. Rev. A
  \textbf{80}, 053606 (2009).

\bibitem{orso} G. Orso, A. Iucci, M. A. Cazalilla and T. Giamarchi, 
Phys. Rev. A \textbf{80}, 033625 (2009).

\bibitem{aa} S. Aubry and G. Andr\'e, Ann. Israel
  Phys. Soc. \textbf{3}, 133 (1980). 

\bibitem{Fisher89} M. P. A. Fisher, P. B. Weichman, G. Grinstein and
  D. S. Fisher, Phys. Rev. B \textbf{40}, 546 (1989). 

\bibitem{harper} G.-L. Ingold, A. Wobst, C. Aulbach, and P. Hanggi,
Eur. Phys. J. B \textbf{30}, 175 (2002).

\bibitem{nielsen} M. A. Nielsen and I. L. Chuang, \emph{Quantum
    Computation and Quantum Information}, Cambridge University Press
  (2000).

\bibitem{lieb} E. H. Lieb, R. Seiringer, and J. Yngvason, Phys. Rev. B
  \textbf{66}, 134529 (2002).

\bibitem{stephan} J.-M. St\'ephan, S. Furukawa, G. Misguich, and
  V. Pasquier, Phys. Rev. B \textbf{80}, 184421 (2009).

\bibitem{milburn} A. P. Hines, R. H. McKenzie and G. J. Milburn,
Phys. Rev. A \textbf{67}, 013609 (2003).

\bibitem{a1} M. Duncan, A. Foerster, J. Links, E. Mattei, N. Oelkers,
A. P. Tonel,  Nucl. Phys. B \textbf{767}, 227 (2007).

\bibitem{a2} G. Santos, A. Foerster, J. Links, E. Mattei and S. R. Dahmen,
Phys. Rev. A \textbf{81}, 063621 (2010).  

\end{thebibliography}
\end{document}